\documentclass{article}
\usepackage{ijcai19}

\usepackage{times}
\usepackage{soul}
\usepackage{url}
\usepackage[hidelinks]{hyperref}
\usepackage[utf8]{inputenc}
\usepackage[small]{caption}
\usepackage{graphicx}
\usepackage{amsmath}
\usepackage{amsfonts,amssymb}
\usepackage{booktabs}
\title{DeepRec: An Open-source Toolkit for Deep Learning based Recommendation\footnote{Accepted by IJCAI-19 Demonstrations Track}}

\author{
Shuai Zhang$^1$\and
Yi Tay$^2$\and
Lina Yao$^1$\and
Bin Wu$^3$\and
Aixin Sun$^2$\\
\affiliations
$^1$University of New South Wales\\
$^2$Nanyang Technological University\\
$^3$Zhengzhou Univeristy
\emails
\{shuai.zhang@student., lina.yao@\}unsw.edu.au, \\
\{ytay017@e., axsun@\}ntu.edu.sg, wubin@gs.zzu.edu.cn
}

\begin{document}

\maketitle

\begin{abstract}
Deep learning based recommender systems have been extensively explored in recent years. However, the large number of models proposed each year poses a big challenge for both researchers and practitioners in reproducing the results for further comparisons. Although a portion of papers provides source code, they adopted different programming languages or different deep learning packages, which also raises the bar in grasping the ideas. To alleviate this problem, we released the open source project: \textbf{DeepRec}. In this toolkit, we have implemented a number of deep learning based recommendation algorithms using Python and the widely used deep learning package - Tensorflow. Three major recommendation scenarios: rating prediction, top-N recommendation (item ranking) and sequential recommendation, were considered. Meanwhile, DeepRec maintains good modularity and extensibility to easily incorporate new models into the framework. It is distributed under the terms of the GNU General Public License. The source code is available at github: \url{https://github.com/cheungdaven/DeepRec}

\end{abstract}

\section{Introduction}
In recent years, deep learning has achieved tremendous success in many application domains, such as computer vision, natural language processing, and speech recognition. Inspired by the effectiveness of deep neural networks, researchers and practitioners started to apply deep learning techniques to recommendation tasks and have achieved a huge improvement on many recommendation tasks. Meanwhile, the number of publications on deep neural network based recommendation models also increases exponentially~\cite{zhang2017deep}. Moreover, many companies resort to deep learning techniques to improve their recommendation quality for better user experience and increasing revenue~\cite{covington2016deep,cheng2016wide}. In general, deep learning techniques can be utilized to learn representations for user and item side information (e.g., user profile, item content, etc.), introduce nonlinearities to recommendation models and model the sequential patterns of user consumption history. In addition, deep neural networks are of high flexibility, which makes it possible to develop more powerful models by combining different neural networks.

However, this hype also poses a big challenge to researchers and practitioners, that it, reproducing results for those proposed deep learning based models. Although some authors provide source code for reproductivity, many of them are written in different programming languages, or use different deep learning packages, let alone a large portion of work do not provide source code, which makes it hard to understand and hinder the process of justifying newly proposed methods. To this end, we start the project - DeepRec. Our goal is to make reproducing results easier and lower the bar of developing deep learning based recommendation models and increase their practicality.

Although there are some open source recommendation packages available, such as MyMediaLite\footnote{http://mymedialite.net/index.html}, LibRec~\cite{guo2015librec}, Suprise\footnote{http://surpriselib.com/} and OpenRec~\cite{yang2018openrec}, most of them only include traditional recommendation algorithms or have not updated for a long time. The OpenRec library do provide several deep neural network based recommendation models, the number is still very small (only four models implemented) compared with DeepRec.

To this end, we develop an open-source toolkit named `\textbf{DeepRec}'. This toolkit provides a flexible framework and unified interfaces for deep learning based recommendation models.  The prime features of DeepRec are twofold:

\begin{itemize}
    \item In this toolkit, we addressed three common scenarios in recommendation research: rating prediction, top-n recommendation (item ranking) and sequence-aware recommendation. It offers state-of-the-art algorithms for those three tasks.
    \item DeepRec is extensible, adaptable and efficient. Users could add algorithm easily. Moreover, DeepRec is able to make use of the computation power of GPUs, making it scalable to large-scale datasets.
\end{itemize}

\section{Introduction to DeepRec}
In this section, we briefly introduce the archecture of the DeepRec, plus implemented recommendation algorithms and comparison with other frameworks.

\subsection{Architecture of DeepRec}
DeepRec is made up of three components: recommender, backend and utility. We adopt the widely used deep learning framework Tensorflow and several other frequent used scientific calculation libraries such as Numpy, Scipy, Sklearn and Pandas as the core backend;  A number of representative and advanced algorithms deep learning based recommendation models forms the recommender components. We split the algorithms into three categories: rating prediction, top-n recommendation and sequence-aware recommendation. We put the common evaluators and data preprocess functions into the utility component.

\begin{figure}
    \centering
    \includegraphics[width=0.48\textwidth]{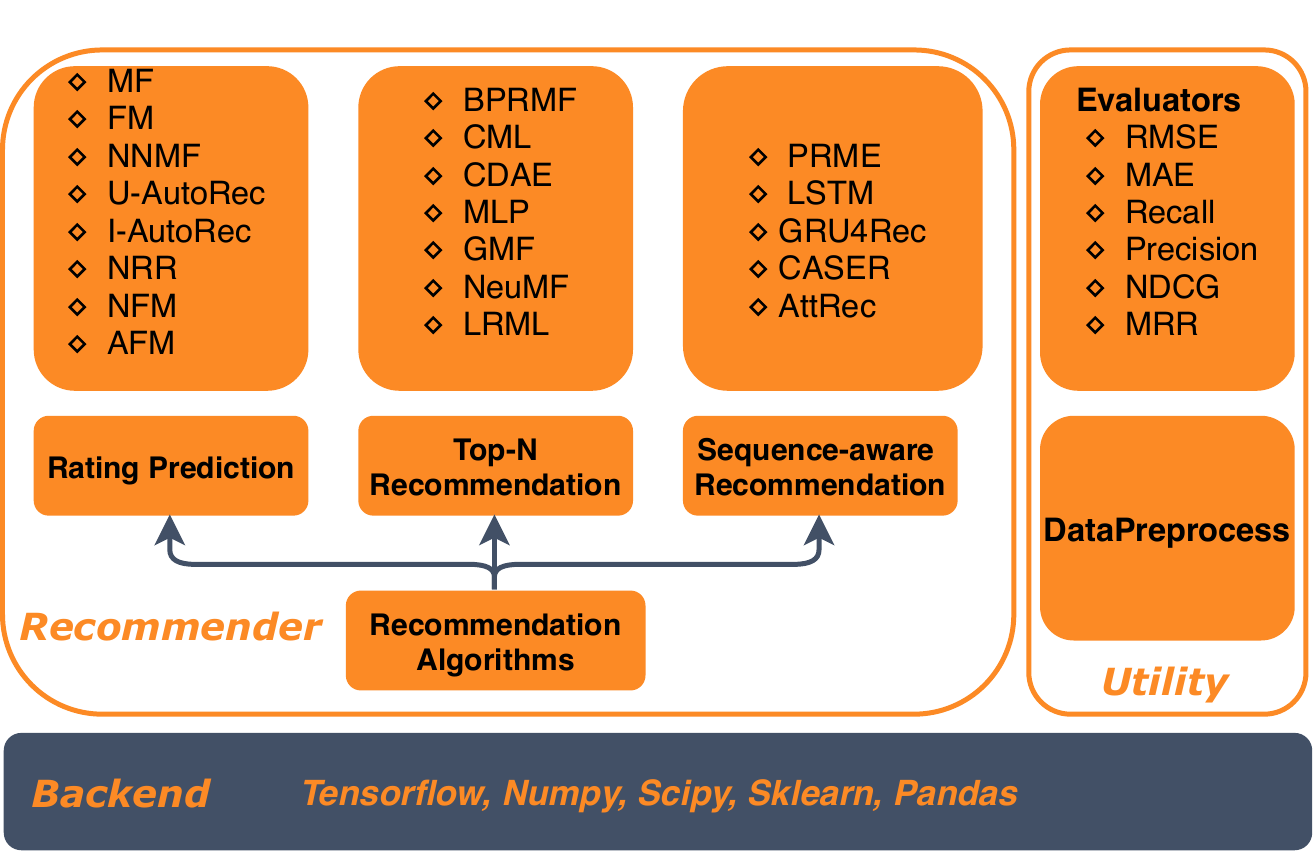}
    \caption{Architecture of DeepRec. It consists of three important components: recommender, backend and utility.}
    \label{fig:selfatt}
\end{figure}

\subsection{Implemented Algorithms} A number of algorithms that were implemented in the current version are listed below. Note that a few non deep learning methods (e.g., biasedSVD, Factorization Machines, BPRMF) are also implemented with Tensorflow.

\subsubsection{Rating Prediction Models} In this category, we implemented methods like classical method like biasedSVD (or MF)~\cite{koren2009matrix}, Factorization Machines (FM)~\cite{rendle2010factorization},  user and item based AutoRec~\cite{sedhain2015autorec}, neural network matrix factorization (NNMF)~\cite{DBLP:journals/corr/DziugaiteR15}, neural rating regression (NRR)~\cite{li2017neural}, neural factorization machines (NFM)~\cite{he2017neuralfm} and attentive factorization machines (AFM)~\cite{xiao2017attentional}. For factorization machines, we use the same feature format as libfm~\cite{rendle2012factorization} to make them compatible.

\subsubsection{Top-n Recommendation Models} Here, we included classical pairwise personalized ranking methods BPRMF~\cite{rendle2009bpr}, collaborative metric learning (CML)~\cite{hsieh2017collaborative}, collaborative denoising auto-Encoder (CDAE)~\cite{wu2016collaborative}, multilayer-perceptron (MLP), generalized matrix factorization (GMF) and neural collaborative filtering (NeuMF)~\cite{he2017neural}, NeuRec~\cite{zhang2018neurec} and latent relational metric learning~\cite{tay2018latent}.

\subsubsection{Sequential Recommendation Models} For sequence-aware recommendation models, we implemented the personalized ranking metric embedding model (PRME)~\cite{feng2015personalized}, LSTM and GRU based methods~\cite{hidasi2015session}, convolution based model (Caser)~\cite{tang2018personalized}, and self-attention based approach (AttRec)~\cite{zhang2018next}.

\subsection{Evaluation Metrics}
Several commonly used evaluation metrics were included in this toolkit. For rating prediction, we can use Root Mean Square Error (RMSE) and Mean Average Error (MAE) to measure the effectiveness. For top-n recommendation and sequence-aware recommendation, we included several ranking-aware metrics, including, Recall@n, Precision@n, Normalized Discounted Cumulative Gain (NDCG) and  Mean Reciprocal Rank (MRR) to measure both the accuracy and quality of ranking lists. For detailed definitions, readers are referred to \cite{shani2011evaluating}.

\subsection{Comparison with Other Frameworks}

\begin{table}[h]
\small
\begin{tabular}{|c|c|c|c|c|}
\hline
\textbf{Library}      & \textbf{backend}    & \textbf{GPUs} &  \textbf{3 scenarios} & \textbf{\#DL}  \\ \hline
Suprise     & SciKits    & N    & N               & 0            \\ \hline
MyMediaLite & N/A        & N    & N               & 0            \\ \hline
LibRec      & N/A        & N    & N               & 0            \\ \hline
OpenRec     & Tensorflow & Y    & N               & 4            \\ \hline
\textbf{DeepRec}     & Tensorflow & Y    & Y               & \textgreater{} 20 \\ \hline
\end{tabular}
\caption{Comparison with existing libraries. ``3 scenarios'' means whether the library can deal with the above mentioned three different tasks. ``\#DL'' means the number of neural networks based methods.}
\label{tab:comparison}
\end{table}

Table \ref{tab:comparison} summarizes the difference between our toolkit and other existing Toolkits. Suprise, MyMedialite, LibRec do not include neural network based models and can not utilize the computation power of GPUs. OpenRec also adopts Tensorflow as the backend, but it only implemented four ranking based recommendation methods, while DeepRec includes over twenty state-of-the-art deep learning based recommendation model. In addition, DeepRec is the first library that includes sequence-aware models.



\section{Conclusion}

In this demo paper, we introduce the open-source library DeepRec, in the hope that it could ease the burden of researchers and practitioners on reproducing existing models. DeepRec is a flexible, extensible and efficient framework which includes a number of state-of-the-art deep learning based recommendation algorithms.


Moving forward, future work will include: implementing more models and evaluation measures, improving the documentation and tutorials as well as updating it to follow the scheme of Tensorflow 2.0.

\bibliographystyle{named}
\bibliography{ijcai19}

\end{document}